\documentclass[final,conference, letterpaper]{IEEEtran}
\IEEEoverridecommandlockouts
\usepackage{cite}
\usepackage{amsmath,amssymb,amsfonts}
\usepackage{algorithmic}
\usepackage{graphicx}
\usepackage{textcomp}
\usepackage{xcolor}

\usepackage{wrapfig}
\usepackage{subcaption}
\usepackage{tabularx}
\usepackage{booktabs}
\usepackage{pifont}
\def\BibTeX{{\rm B\kern-.05em{\sc i\kern-.025em b}\kern-.08em
    T\kern-.1667em\lower.7ex\hbox{E}\kern-.125emX}}
\usepackage[top=0.75in, bottom=1.08in, left=0.625in, right=0.64in, columnsep=0.28in]{geometry}
\usepackage{microtype}
\let\oldbibliography\thebibliography
\renewcommand{\thebibliography}[1]{%
  \oldbibliography{#1}%
  \setlength{\itemsep}{0pt}%
  \setlength{\parskip}{0pt}%
}

\begin{document}

\title{Vision-based Sensing of Urban Environment using Hardware-Software co-design at Scale
}
\author{\IEEEauthorblockN{Akshit Gupta, R Venkatesha Prasad}
\IEEEauthorblockA{
\textit{TU Delft},  
Delft, The Netherlands
a.gupta-5@tudelft.nl; r.r.venkateshaprasad@tudelft.nl
\thanks{This work has been accepted for publication in the IEEE World Forum on Internet of Things (WF-IoT). © 2026 IEEE. Personal use of this material is permitted. Permission from IEEE must be obtained for all other uses, in any current or future media, including reprinting/republishing this material for advertising or promotional purposes, creating new collective works, for resale or redistribution to servers or lists, or reuse of any copyrighted component of this work in other works.}}
}

\maketitle

\begin{abstract}
Urban environmental monitoring is crucial for climate-adaptive cities, yet current approaches rely heavily on human inspection, overhead remote sensing, or limited terrestrial RGB imagery, thereby lacking comprehensive multispectral perspectives at street level. To bridge this gap, this paper introduces UrbanScapes, an adaptable and flexible hardware-software co-design platform for large-scale, vision-based sensing of urban environments. Built on a modular architecture, UrbanScapes seamlessly integrates multispectral imaging channels, including Red, Green, Blue (RGB), Near-Infrared, and Thermal, to capture multispectral street-level imagery. The system is designed for both opportunistic deployment on bicycles, vehicles, or pedestrians and targeted deployments on municipal fleets, ensuring high spatio-temporal scalability without requiring hardware redesigns. Crucially, the platform utilizes on-device deep learning models to automatically remove personally identifiable information (PII) at the edge, addressing privacy concerns inherent in street-level data collection. By combining this multispectral street-level imagery with overhead multispectral remote sensing imagery, UrbanScapes enables scalable calculation of physics-based environmental indices, providing a foundational tool for advanced citywide analysis.


\end{abstract}


\section{Introduction}
Urban environmental monitoring is vital for the development of climate-resilient cities. 
Existing approaches, largely comprising remote sensing imagery and manual human inspections, are limited to overhead views and non-scalable, respectively. In this work, we demonstrate an integrated hardware-software platform, UrbanScapes, for opportunistic sensing to collect multispectral street level data. This data can be combined with overhead remote sensing imagery, enabling comprehensive monitoring of urban environmental parameters at city-wide scales. Specifically, the UrbanScapes hardware comprises Red, Green, and Blue (RGB), near-infrared, and thermal imaging sensors that collect multispectral street-level imagery synchronously at high spatial and temporal resolutions. Data quality is ensured by utilizing similar imaging channels available in overhead remote sensing imagery, which is later processed using computer vision-based deep learning models to understand and quantify various urban environmental parameters. For instance, in the urban city context, one can identify different building façade materials using near-infrared channels, analyze urban greenery using a combination of vision and near-infrared channels or analyze energy leakage from building façades using thermal channels. The device is designed for flexible deployment via magnets on bikes, cars, and buses, and can also be carried by humans. Thus, it is suitable both for participatory citizen-science based sensing and opportunistic municipal sensing. Further, all personally identifiable information (PII) is removed on-device using a deep learning model. Hence, enabling comprehensive analysis of urban environmental parameters at scale supported by explainable physics-based indices. In this paper, we highlight the design of the system, architecture of the system, and demonstrate capabilities possible from multispectral street-level data collected by this device. The device was deployed in real-world across 4 urban areas inside the Netherlands collecting synchronized multi-spectral imagery, generating a dataset of 17k street-level images, and available openly as the Spectrascapes dataset \cite{SpectrascapesDataset}.

\section{Related Work}
Overhead remote sensing imagery (using drones, satellites, and airplanes) generates massive amounts of data across multiple spectral channels, in addition to the standard Red, Green, and Blue (RGB) channels visible to the human eye. These additional spectral channels are useful since they allow the calculation of quantitative and calibrated physics-based environmental indices such as Normalized Difference Vegetation Index (NDVI). However, most studies in the last decade, such as \cite{tarkhan2025mapping,2024_global_streetscapes}, are often limited by terrestrial RGB imagery from platforms such as Google Street View (GSV), KartaView, and Baidu Maps \cite{biljecki_2021_street}.Recognizing the value of additional spectral channels in terrestrial or street-level imagery, some recent projects have deployed custom hardware to capture spectral channels beyond the standard Red, Green, and Blue. For example, in \cite{instantInfrared}, thermal and RGB imaging sensors are utilized to detect energy leakage from building facades, while in \cite{greenscan}, thermal and near-infrared imaging sensors are utilized on moving vehicles to detect the health condition of trees. However, these works focus on selecting specific spectral bands based on the application domain, thereby, limiting the flexibility of their hardware platform to capture additional spectral channels. Furthermore, in these studies, the designed system does not account for the removal of personally identifiable information (PII). Table \ref{tab:comparison} summarizes these works from several aspects. 
\begin{table}[h]
\centering
\resizebox{\columnwidth}{!}{%
\begin{tabular}{lcccccc}
\toprule
Work & Custom Platform & \multicolumn{3}{c}{Spectral Channels in Imagery} & Platform Flexibility & PII removal\\
\cmidrule(lr){3-5} & & RGB & NIR & Thermal & &  \\
\midrule
\cite{tarkhan2025mapping}  & No & \ding{51} & \ding{55} & \ding{55}  & Zero & Yes \\
\cite{2024_global_streetscapes} & No & \ding{51} & \ding{55} & \ding{55}  & Zero & Yes \\
\cite{greenscan}    & Yes & \ding{51} & \ding{51} & \ding{55}  & Low & No  \\
\cite{instantInfrared} & Yes    & \ding{51} & \ding{55} & \ding{51}  & Low & No  \\
\textbf{This work}  & Yes & \ding{51} & \ding{51} & \ding{51}  & High &  Yes \\
\bottomrule
\end{tabular}%
}
\caption{Comparison of recent works in street-level view imagery domains in terms of spectral richness, flexibility, PII removal and our contributions.}
\label{tab:comparison}
\end{table}

\section{Research contributions}
\begin{figure}[hbt!]
\centering
\includegraphics[width=\linewidth]{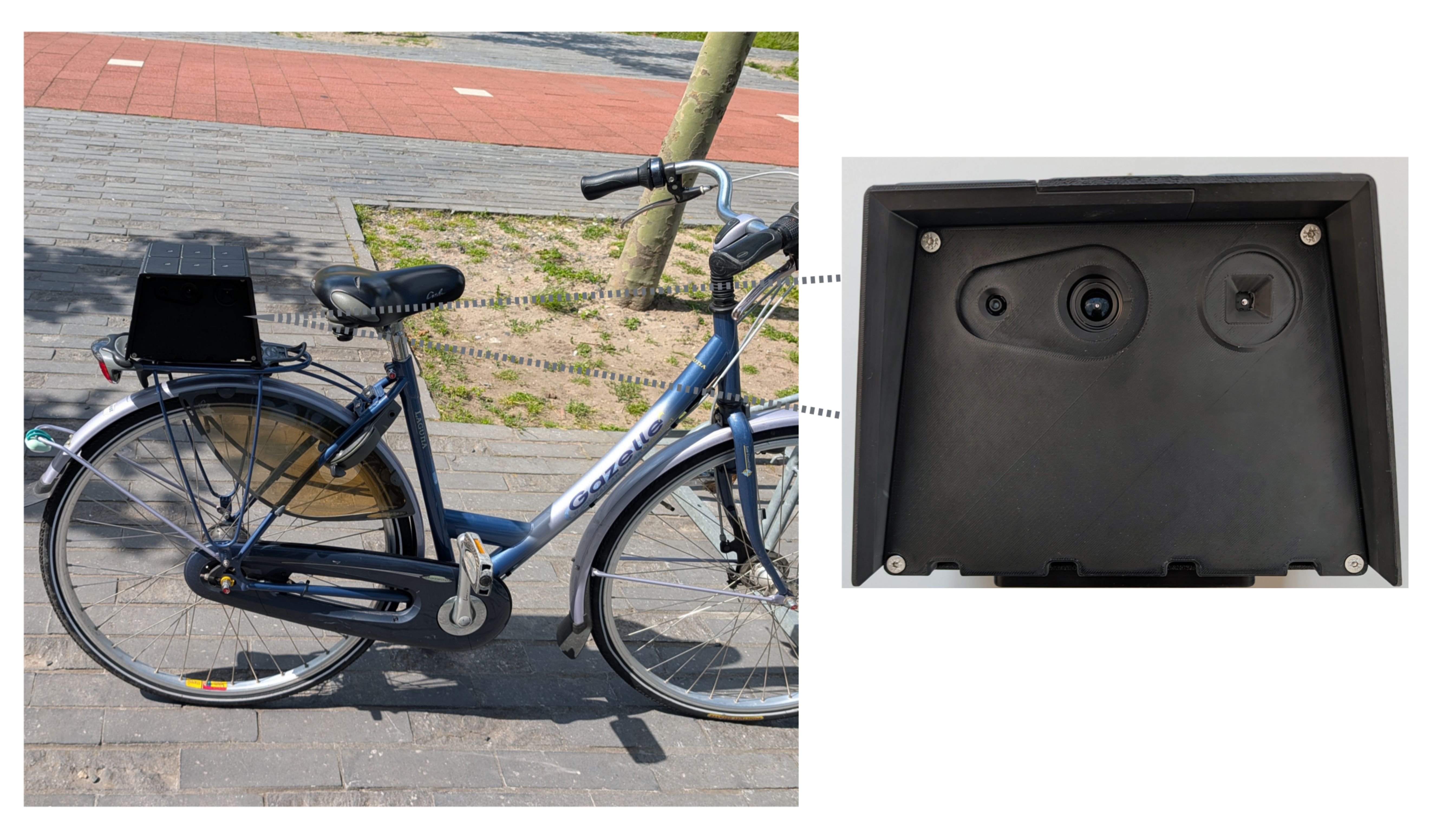}
\caption{The UrbanScapes device with three imaging sensors and its deployment on a bicycle with magnets.}
\label{fig:deployment}
\end{figure}

Our platform is shown in Figure \ref{fig:deployment}. Addressing the research gaps (Table \ref{tab:comparison}), our approach has the following research contributions:
\begin{itemize}
    \item \textbf{Platform flexibility and extension:} The architecture of Urbanscapes is designed around a central single board computer based on a Linux kernel. Utilizing a single-board computer enables the integration of diverse hardware interfaces and software architectures. This allows flexibility in both port (or interface) utilization and software module development for each imaging sensor, with no major redesign required, such as soldering new components. In our current configuration, a Raspberry Pi 4 was utilized with the three imaging sensors connected and controlled via UART, SPI, and I2C interfaces.
    \item \textbf{Ease of Deployment:} The platform is designed for deployment on bicycles, vehicles (such as buses, cars) with magnets, or to be carried by humans. Thus, allowing monitoring urban parameters using opportunistic municipal sensing fleets and micro-mobility sensing on bikes or citizen-science based initiatives. This deployment structure enables scalability in both spatial and temporal coverage.
    \item \textbf{Data Quality and Privacy}: Our street-level imagery can be validated and calibrated using existing remote sensing data, since our system uses the same spectral channels available in satellite data. Thus ensuring data quality. To ensure PII removal, Ultralytics YOLO object detection framework \cite{Jocher_Ultralytics_YOLO26_Unified_2026} is utilized after a specific batch of images is captured. This ensures the detection and gaussian blurring of human faces and vehicle license plates on the device itself.
\end{itemize}
%
%
\begin{figure}[hbt!]
\centering
\includegraphics[width=\linewidth]{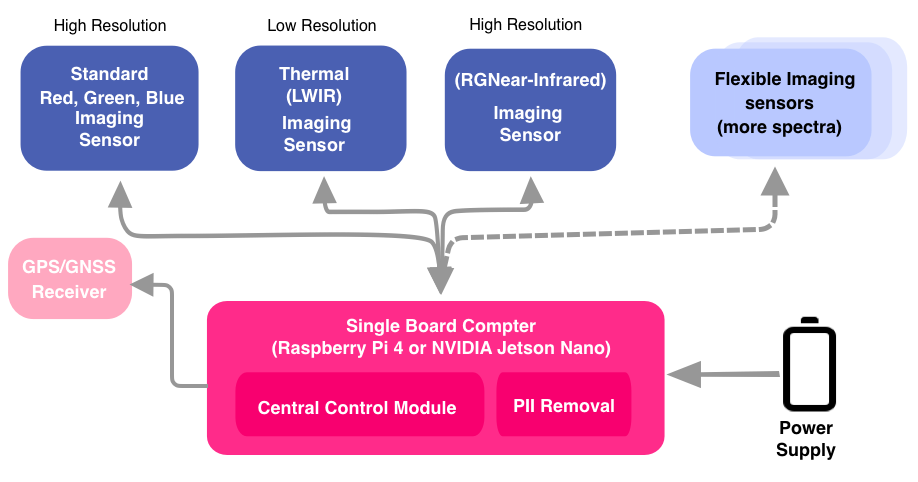}
\caption{The architecture of the system. Imaging data is geo-tagged with a provision for up to six imaging sensors with different spectral channels.}
\label{fig:architecture}
\end{figure}

\section{System Design and Architecture}
\begin{figure}[hbt!]
\centering
\includegraphics[width=\linewidth]{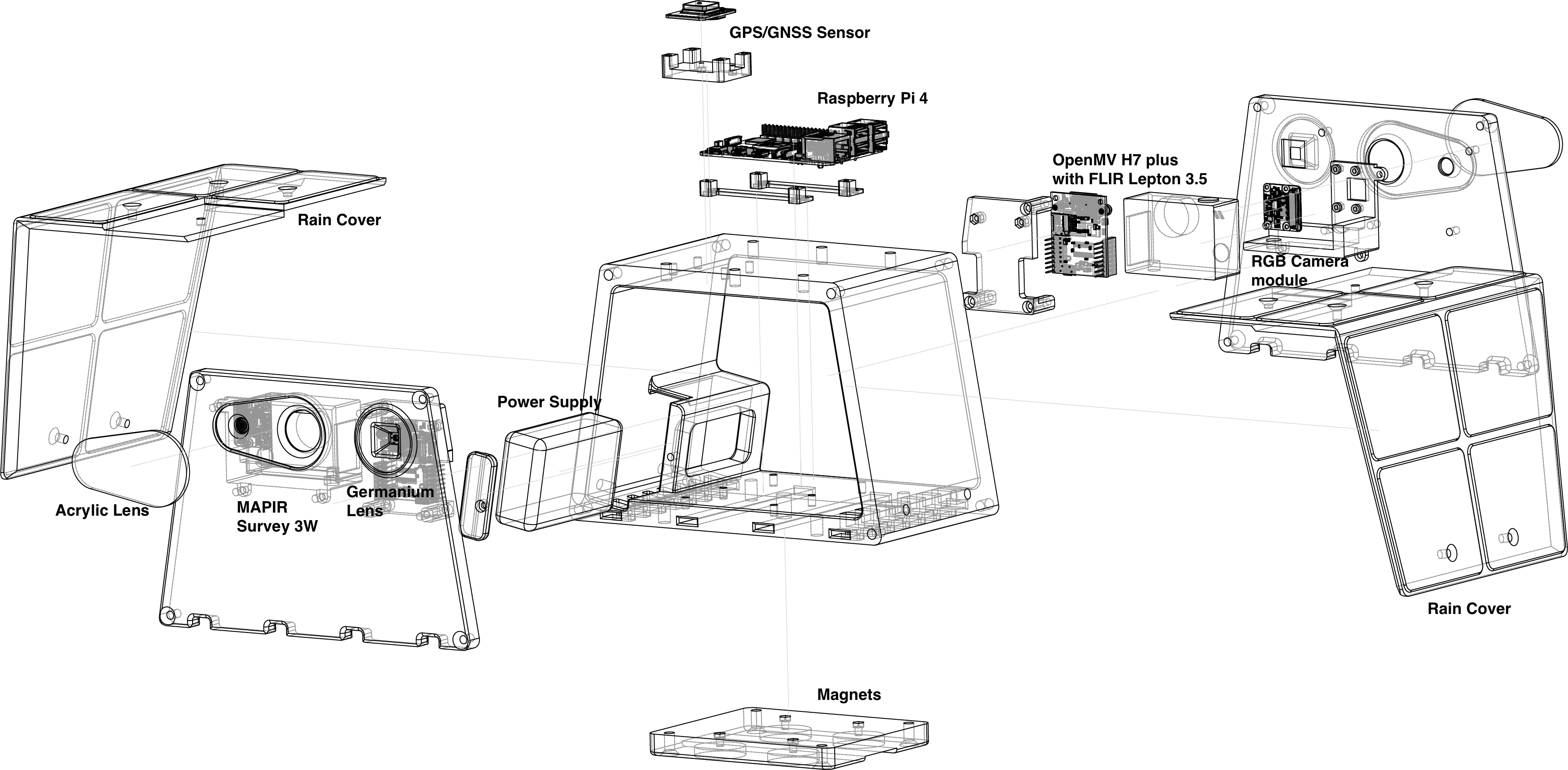}
\caption{The assembly of all the components. The design allows imaging sensors to be installed on both sides of the device.}
\label{fig:assembly}
\end{figure}

The architecture of UrbanScapes is shown in Figure \ref{fig:architecture}.
The exploded-view diagram showing the assembly of all components used for different spectral channels is shown in Figure \ref{fig:assembly}. 
In its current state, the casing design allows a dual-sided view and supports six imaging sensors. The casing is 3D-printed from ASA (Acrylonitrile Styrene Acrylonitrile) filament, making it weather resistance at a low-cost. The device measures $26\text{ cm} \times 14\text{ cm} \times 15\text{ cm}$ and weighs $1.25\text{ kg}$, including all hardware components and three imaging sensors.

%
\section{Deployment and results}
An early prototype of the Urbanscapes device was deployed in real-world urban campaigns across four different urban morphologies in the Netherlands (Arnhem, Delft, Oirschot, and Middelbeers), representing large urban, university/medium urban, town, and rural village settings, respectively. The details of deployment, the dataset gathered, and some early use-cases are also available at \cite{SpectrascapesDataset}.
\newline

\textbf{Field Deployment and System Performance} \newline
The system traversed a cumulative distance of 82~km mounted on bicycles over a 7-day period. Bicycles allow agile navigation through narrow historical alleyways and dedicated bike paths, maintaining close proximity to building façades and urban canopy layers. In total, 17,718 synchronized multi-spectral image triplets were captured across three optical modalities: RGB ($4608\times 2592$), RGN ($4000\times 3000$), and Thermal LWIR ($160\times 120$).

Operating on a standard $10,000\text{ mAh}$ power pack, the device consumed under $7.8\text{ W}$, yielding $\sim 4.5\text{ hours}$ of continuous mobile operation. At typical cycling speeds ($8-10\text{ km/h}$), each frame came with metadata comprising  hardware-synchronized timestamps and GPS telemetry data, without blur.
\newline

\begin{figure}[hbt!]
\centering
\includegraphics[width=0.8\linewidth]{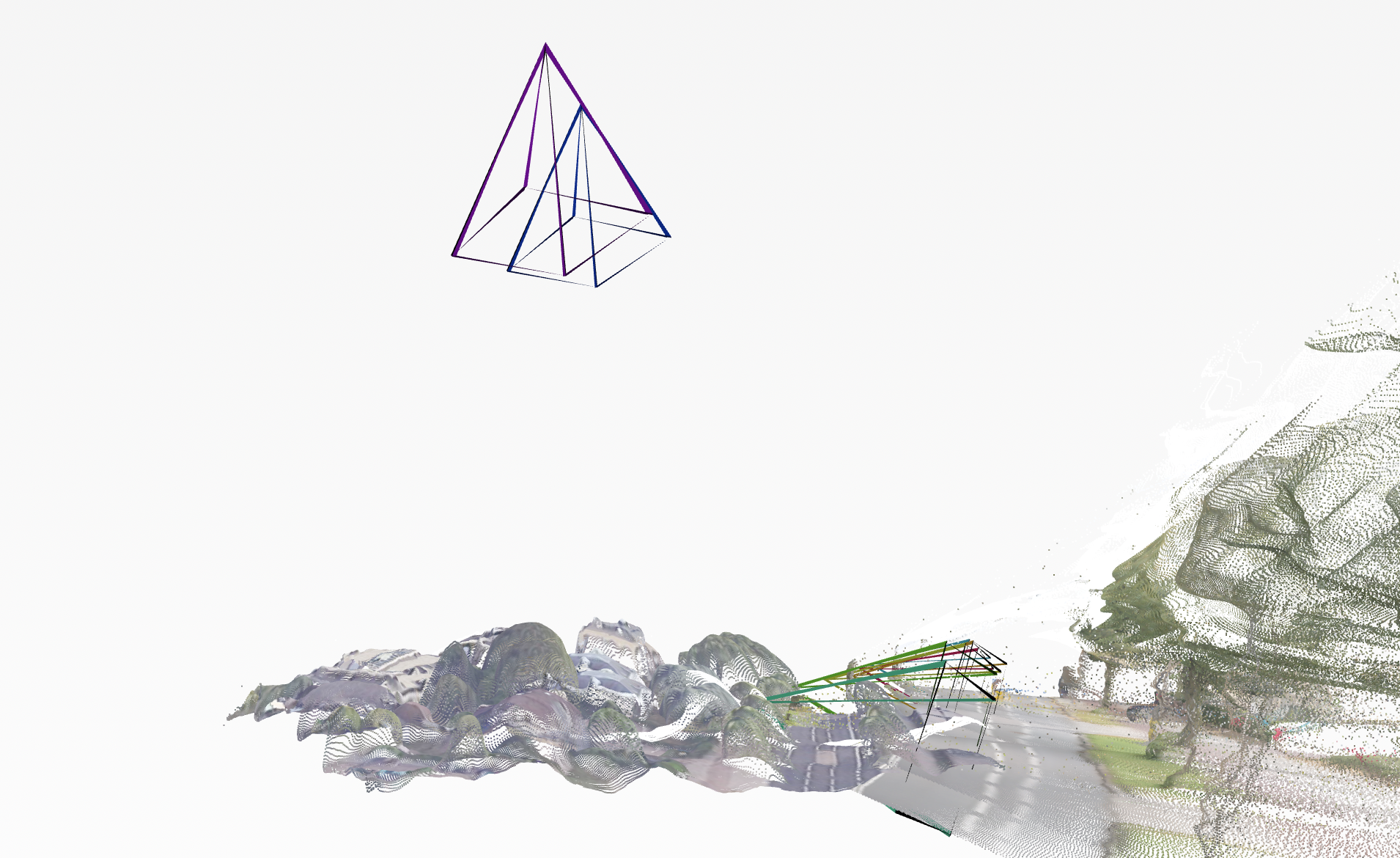}
\caption{Comprehensive 3D reconstruction using street-level and remote sensing imagery for a street in Arnhem, The Netherlands. Purple and Blue cones denote airborne imagery data obtained from Maxar Technologies (accessed from Google Maps) and the remaining cones denote street-level imagery data captured using UrbanScapes.}
\label{fig:reconstruction}
\end{figure}

\textbf{Cross-Scale 3D Reconstruction and Multi-Spectral Utility}
Using this as a new modality in state-of-the-art deep learning models for 3D reconstruction such as ZipMap \cite{jin2026zipmap},
we combined terrestrial images captured using this platform with overhead airborne imagery. An example of this reconstruction is shown in Figure \ref{fig:reconstruction}, for a street in Arnhem, The Netherlands. As seen in the figure, this type of reconstruction enables a comprehensive terrestrial and overhead digital twin of the built environment, with multispectral channels enabling calculation of physics-based environmental indices. With the dataset captured using field deployment, one can perform this 3D reconstruction across the entire trajectory of the bicyle movement.

\section{Conclusion and Future}
We showcased a hardware-software co-design platform called UrbanScapes to gather multispectral street-level data with scalability. The architecture is designed with flexibility of sensors and deployments. The integration of this multispectral imagery data with remote sensing imagery data enables calculation of physics-based environmental indices. The deployment modes can be either opportunistic municipal fleets or micro-mobility sensinf with bicycles or citizen science intitiatives. The dual-side ASA housing with synchronized image capture and removal of PII makes it suitable for public space image capture. The device was deployed in real-world across 4 cities in the Netherlands generating a dataset with 17,718 synchronized images. Future work includes deployement suited for temporal data collection and analysis using public buses in the Netherlands.

\bibliographystyle{IEEEtran}
\bibliography{references}

\end{document}